\documentclass[aps,preprint,epsfig,superscriptaddress]{revtex4-1}

\usepackage{amsmath,amssymb,graphicx}
\usepackage{latexsym}
\usepackage{esint}
\usepackage{color}
\usepackage{float}
\usepackage{multirow}
\usepackage{hyperref}
\usepackage{enumerate}
\usepackage{amsbsy}
\usepackage{bm}

\newcommand{\be}{\begin{equation}}
\newcommand{\ee}{\end{equation}}
\newcommand{\bea}{\begin{eqnarray}}
\newcommand{\eea}{\end{eqnarray}}

\newcommand{\comment}[1]{}

\begin{document}

\title{Ordering Dynamics of the Random Field Long-range Ising Model in One Dimension}

\author{Ramgopal Agrawal}
\email{ramgopal.sps@gmail.com}
\email{ramgopal@lpthe.jussieu.fr}
\affiliation{School of Physical Sciences, Jawaharlal Nehru University, New Delhi 110067, India.}
\affiliation{Sorbonne Universit\'e, Laboratoire de Physique Th\'eorique et Hautes Energies, CNRS UMR 7589, 4 Place Jussieu, 75252 Paris Cedex 05, France.}
\author{Federico Corberi}
\email{corberi@sa.infn.it}
\affiliation{Dipartimento di Fisica ``E.~R. Caianiello'', and INFN, Gruppo Collegato di Salerno, and CNISM, Unit\`a di Salerno, Universit\`a  di Salerno, via Giovanni Paolo II 132, 84084 Fisciano (SA), Italy.}
\author{Eugenio Lippiello}
\email{eugenio.lippiello@unicampania.it}
\affiliation{Dipartimento di Matematica e Fisica, Universit\`a della Campania, Viale Lincoln 5, 81100, Caserta, Italy.}
\author{Sanjay Puri}
\email{purijnu@gmail.com}
\affiliation{School of Physical Sciences, Jawaharlal Nehru University, New Delhi 110067, India.}

\begin{abstract}
We investigate the influence of long-range (LR) interactions on the phase ordering dynamics of the one-dimensional random field Ising model (RFIM). Unlike the usual RFIM, a spin interacts with all other spins through a ferromagnetic coupling that decays as $r^{-(1+\sigma)}$, where $r$ is the distance between two spins. In the absence of LR interactions, the size of coarsening domains $R(t)$ exhibits a crossover from pure system behavior $R(t) \sim t^{1/2}$ to an asymptotic regime characterized by logarithmic growth: $R(t) \sim (\ln t)^2$. The LR interactions affect the pre-asymptotic regime, which now exhibits ballistic growth $R(t) \sim t$, followed by $\sigma$-dependent growth $R(t) \sim t^{1/(1+\sigma)}$. Additionally, the LR interactions also affect the asymptotic logarithmic growth, which becomes $R(t) \sim (\ln t)^{\alpha(\sigma)}$ with $\alpha(\sigma) < 2$. Thus, LR interactions lead to faster growth than for the nearest-neighbor system at short times. Unexpectedly, this driving force causes a slowing-down of the dynamics ($\alpha < 2$) in the asymptotic logarithmic regime. This is explained in terms of a non-trivial competition between the pinning force caused by the random field and the driving force introduced by LR interactions. We also study the spatial correlation function and the autocorrelation function of the magnetization field. The former exhibits superuniversality for all $\sigma$, i.e., a scaling function that is independent of the disorder strength. The same holds for the autocorrelation function when $\sigma<1$, whereas a signature of the violation of superuniversality is seen for $\sigma>1$.
\end{abstract}

\maketitle

\section{Introduction}
\label{s1}

Consider a ferromagnetic spin system which is quenched from the paramagnetic phase at temperature $T > T_c$ to $T < T_c$. In the subsequent evolution, spatial correlations build up via the growth of domains, a process known as \textit{phase ordering dynamics} or \textit{coarsening}~\cite{PuriWad09,dp04,Bray94}. This process is controlled by the motion of interfaces~\cite{CorLipZan08} in the system. The additional presence of quenched (static) random disorder~\cite{doi:10.1142/3517,CorCugYos11,Puri2004,PhysRevLett.54.2708,PhysRevE.65.046114,corberi2011growth,corberi2012crossover}, e.g., on-site random fields, pins these interfaces in local free energy minima. Thus, free energy barriers are created which the system must overcome in order to activate the growth process. Many random systems involve long-range (LR) interactions~\cite{campa2014physics}, which play an important role in the static and dynamic properties. A pure system with ferromagnetic LR interactions exhibits coarsening phenomena akin to a short-range system, although the domain growth kinetics may be faster due to cooperativity between far-away spins. Over the past few years, there has been significant progress~\cite{corberi2019one,PhysRevE.102.020102,CMJ19,PhysRevLett.125.180601,PhysRevE.103.012108,PhysRevE.103.052122,PhysRevE.105.034131,Corberi_2019} in understanding the nature of ordering dynamics in disorder-free LR systems. However, very little is known about random systems with LR interactions. This is the subject of the present paper. For simplicity, we will focus here on one-dimensional ($1D$) LR Ising systems. The reasons for this are as follows: \\
(a) The corresponding pure case is well understood and can be used as a reference point to understand the changes introduced by the quenched disorder; and \\
(b) The $1D$ system is a good starting point to investigate this challenging problem since numerical simulations are more reliable. There is a hope of developing some analytical insights, similarly to what was done without disorder.

{\bf Equilibrium aspects --} The pure $1D$ Ising model with nearest neighbor (NN) interactions does not exhibit LR order (LRO) at any temperature other than $T = 0$. The inclusion of a quenched random on-site field yields the well-known {\it random field Ising model} (RFIM). From the Imry-Ma argument~\cite{PhysRevLett.35.1399}, there exists no LRO at any $T$ (including $T=0$) in the RFIM with NN interactions in $D \le 2$. It is likely that, due to the random pinning field at each site, a spin may flip in accordance with the field strength. However, for strong ferromagnetic coupling, the spins succeed to form large correlated domains in the ground state at $T=0$, known as Imry-Ma domains. The linear size of these domains ($R_{\rm IM}$) depends on the ratio between the ferromagnetic constant ($J_0$) and the random field strength ($\Delta$)~\cite{PhysRevLett.35.1399}:
\be
\label{eq1}
R_{\rm IM} = \frac{4 J_0^2}{\Delta^2}.
\ee

The situation is rather different for systems with LR interactions. In the pure $1D$ LR Ising model (LRIM), where the spins have a power-law ferromagnetic interaction [$J(r) \propto J_0/r^{1+\sigma}$, $r$ being the distance], Dyson~\cite{dyson1969existence} proved that LRO is possible even at small non-zero $T$ for $0 < \sigma < 1$. For $\sigma \ge 1$, there is LRO only at $T=0$. Leuzzi-Parisi \cite{PhysRevB.88.224204} and Dewenter-Hartmann \cite{PhysRevB.90.014207} recently studied the $1D$ LRIM in the presence of random field disorder, which is referred to as the \textit{random field long-range Ising model} (RFLRIM). They reported a threshold value of $\sigma$ ($=0.5$) below which there is an LRO state at $T=0$ for $\Delta < \Delta_c \simeq 2.39$. For $\sigma > 0.5$, the Imry-Ma argument holds as usual.

{\bf Nonequilibrium aspects --} During the post-quench dynamics of the spin systems, the onset and growth of domains is a gradient-descent mechanism during which the system minimizes its free energy. The domain size $R(t)$ is the characteristic length scale in the system. The growth law for a quench to $T = 0$ of the $1D$ pure NN Ising model is given by $R(t) \sim t^{1/2}$. Since the early-time dynamics of the system is governed by the $T=0$ fixed point, the same law also holds at small non-zero $T$ before equilibration takes place.

The story is similar (but more complex) for $1D$ pure systems with power-law interactions~\cite{corberi2019one}. For a quench to $T=0$, the growth is ballistic: $R(t) \sim t$ for all $\sigma > 0$. However, for non-zero-$T$ quenches, the growth is ballistic only at pre-asymptotic time scales. If $0 < \sigma < 1$, the ballistic growth is followed by a slower power law with a $\sigma$-dependent exponent, also known as the Bray-Rutenberg (BR) \cite{PhysRevE.49.R27,PhysRevE.50.1900} regime, $R(t) \sim t^{1/(1+\sigma)}$. On the other hand, if $\sigma > 1$, this BR regime occurs at intermediate time scales, and is followed by an asymptotic regime with $R(t) \sim t^{1/2}$, akin to the NN case.

In this paper, we investigate the ordering dynamics of the $1D$ RFLRIM with \textit{nonconserved} or {\it Glauber spin-flip} kinetics. This means that single spins can be flipped with a transition rate which obeys detailed balance. We focus on the case where both the random field disorder and the quench temperature are smaller than the ferromagnetic coupling constant, i.e., $\Delta < J_0$ and $T < J_0$. In the absence of LR interactions, the ordering dynamics is quite well understood. Fisher et al.~\cite{PhysRevE.64.066107} found that an interface in the $1D$ RFIM can be envisioned as a random walker in a random force field of the Sinai type~\cite{doi:10.1137/1127028,BOUCHAUD1990285}. They showed via renormalization group techniques that the domain growth law is $R(t) \sim \left( \ln t \right)^2$, which was later numerically verified by Corberi et al.~\cite{PhysRevE.65.046114}. In the Huse-Henley framework~\cite{PhysRevLett.54.2708}, this logarithmic growth is linked to the $R$-dependence of the free energy barrier height.

The presence of LR couplings substantially changes the above scenario. One major difference is that the interfaces now feel a drift due to the LR interactions. This drift tries to balance both the randomization due to thermal noise and the metastability due to the quenched pinning field. To the best of our knowledge, this problem has never been studied in the literature, and we address it in this paper. We consider only the \textit{weak LR limit} ($J(r) \propto 1/r^{1+\sigma}, \sigma > 0$) here. We remind the reader that the \textit{strong LR limit} $(\sigma \le 0)$ is a completely distinct problem, as it causes the loss of additivity and extensivity~\cite{campa2014physics,CorIanKumLipPol21}.

We will address several issues regarding domain growth of the $1D$ RFLRIM. In particular, we study the growth of $R(t)$ which displays a rich crossover form. We also analyze the dynamical scaling properties of the space and time correlation functions of the magnetization field. In random systems, it is relevant to ask whether the scaling function has an explicit dependence on the disorder amplitude. If there is no disorder-dependence other than through the length scale $R(t,\Delta)$, the scaling functions are referred to as \textit{superuniversal} (SU) \cite{pcp91,pp92}. In our framework, the term superuniversality implies that the entire scaling function remains unaffected by variations in the amplitude of disorder. The spatial correlation function usually exhibits SU. However, the validity of SU has been controversial in two-time quantities, e.g., $D=1,2$ RFIM with NN interaction~\cite{PhysRevE.65.046114,corberi2012crossover}, and $D=2,3$ random field XY model~\cite{PhysRevE.104.044123} show clear violations of SU in the autocorrelation function. We examine the applicability of SU for the RFLRIM in great detail in this paper.

Our important observations can be summarized as follows: \\
(a) The characteristic length scale $R(t)$ has a logarithmic time-dependence at asymptotic times. The growth law is slower than that for the NN case. \\
(b) The scaling function for the spatial correlation function $C(r,t)$ shows SU for a given value of $\sigma$. \\
(c) The autocorrelation function exhibits scaling for all $\sigma$. The scaling function exhibits SU for $\sigma < 1$, while it is seemingly violated for $\sigma > 1$.

This paper is structured as follows. In Sec.~\ref{s2}, we introduce the model and simulation details. The important observable quantities which we measure are also discussed here. In Sec.~\ref{s3}, we present detailed numerical results. In Sec.~\ref{s4}, we summarize the results obtained in this paper, and discuss possible future directions.

\section{Model and Simulation Details}
\label{s2}

We consider a $1D$ RFLRIM with the Hamiltonian:
\be
{\cal H}(\{s_i\})= -\sum_{j<i} J(r_{ij}) s_i s_j - \sum _{i=1}^{N} h_i s_i, \quad s_i = \pm 1,
\label{ham}
\ee
where $N$ is the number of lattice sites, i.e., system size. The ferromagnetic exchange coupling between the spins $s_i$ and $s_j$ lying at distance $r_{ij} =\vert i - j \vert$ has the power-law form:
\be
J(r) = \frac{J_0}{r^{1+\sigma}}, \quad \sigma >0 .
\label{int}
\ee
The variables $h_i$ denote quenched random fields drawn from a Gaussian distribution:
\be
P(h_i) dh_i = \frac{1}{\sqrt{2 \pi \Delta^2}} e^{-h_i^2/(2\Delta^2)} dh_i .
\ee
Here, the width of the distribution $\Delta$ controls the disorder strength. One can see in Eq.~\eqref{ham} that the usual NN Ising model is recovered on setting $J(r) = J_0 \delta_{r,1}$. The constant $J_0$ is the strength of the exchange term, and is subsequently set to unity, without any loss of generality.

We consider Glauber spin-flip kinetics \cite{PuriWad09}, where single randomly-chosen spins flip with Metropolis transition rates
\be
w(s_i\to -s_i) = N^{-1}\min \left( 1,e^{-\Delta E/T} \right) .
\label{metrop}
\ee
Here, $\Delta E$ is the energy difference in the proposed move, and we have set the Boltzmann constant to unity. Time is measured in Monte Carlo steps (MCS), each corresponding to $N$ attempted elementary moves. We study the evolution of paramagnetic system in equilibrium at $T = \infty$, which is suddenly quenched to a low temperature at time $t=0$. The system is then evolved with the Metropolis transition rate in Eq.~(\ref{metrop}).

To implement periodic boundary conditions, we exploit a sophisticated approach developed for LR systems~\cite{fs2002,PhysRevE.103.012108}. Consider an infinite $1D$ lattice partitioned into imaginary copies of the original simulation lattice. The central cell is the simulation lattice itself, and the images lie across its periodic boundary. The effective interaction between two spins $s_i$ and $s_j$ inside the simulation lattice can be expressed as an infinite summation over all images:
\be
\label{jsum}
J^*(r_{ij}) = \sum_{n} \frac{1}{\vert n N + i - j \vert ^{1+\sigma}} .
\ee
Here, the displacement index $n = 0, \pm 1, \pm 2, \ldots $, denotes the image systems. The advantage in $1D$ is that the infinite summation in Eq.~\eqref{jsum} can be analytically calculated~\cite{PhysRevB.86.014431} without any cut-off using Hurwitz zeta functions~\cite{abramowitz1988handbook}:
\be
\label{hurz}
H(a,b) = \sum_{m=0}^{\infty} \frac{1}{(b+m)^a}; \quad b \ne 0,~\mbox{Re}(a) > 1.
\ee
With the help of Eq.~\eqref{hurz}, the effective interaction $J^*(r)$ has the form
\be
\label{jsum_2}
J^*(r) = \frac{1}{N^{1+\sigma}} \left[ H\left( 1+\sigma, \frac{r}{N} \right) + H\left( 1+\sigma, 1 - \frac{r}{N} \right) \right].
\ee
To compute Hurwitz zeta functions in our numerical simulations, we have implemented the \textsc{GSL} library~\cite{gough2009gnu} in \textsc{gfortran}.

We define the defect density $\rho(t)$ as the number of misaligned spins divided by $N$. The excess defect density at time $t$ is $\rho_{ex}(t)=\rho(t)-\rho(t=\infty)$, i.e., the difference from the equilibrium value. This quantity is used to evaluate the domain size $R(t) = \overline{\langle\rho_{ex} (t)\rangle}^{-1}$~\cite{Bray94}. Here, $\overline{\left< .... \right>}$ represents an average over independent initial configurations, thermal histories, and disorder realizations. We remark that there are several definitions for calculating the coarsening length scale, or the characteristic domain size. For example, the distance at which the spatial correlation function $C(r,t)$ (defined below) decays to a fixed value provides a measure of the length scale. The ratio of integrals $\int dr r^2 C(r,t)$ and $\int dr r C(r,t)$ provides another way of calculating the coarsening length scale. In the scaling regime, these definitions differ from each other by only a prefactor (see Appendix~\ref{corr_len}). In this paper, we usually compute the length scale from the inverse defect density.

To study the domain growth law quantitatively, we calculate the {\it effective} dynamic exponent $z_{\rm eff}$ as follows:
\be
\label{growth_exp}
z_{\rm eff} (t, \Delta)=\left [\frac{d \ln R (t)}{d \ln t}\right]^{-1}. 
\ee

We also compute the equal-time spatial correlation function $C(r,t)$, defined as
\be
\label{eq5}
C(r ,t) = \frac{1}{N} \sum_{i=1}^N \left[\overline{\left<s_{i}(t) s_{i+r}(t)\right>} - \overline{\left<s_{i}(t)\right>} ~ \overline{\left<s_{i+r}(t)\right>}\right].
\ee
In coarsening systems, this quantity usually exhibits dynamical scaling
\cite{PuriWad09}, i.e.,
\be
\label{cf_scale}
C(r,t) = f\left(\frac{r}{R(t)}\right).
\ee
Here, $f(x)$ is a scaling function. An experimentally measurable quantity is the {\it structure factor} $S(k,t)$, which is the Fourier transform of $C(r,t)$:
\be
\label{sf}
S(k,t) = \sum _r ~{\rm e}^{ikr} C(r,t).
\ee
This quantity has the scaling form $S(k,t) = R(t) f_k\left(k R(t)\right)$, where $f_k(p)$ is another scaling function. In pure systems, due to sharp interfaces, the function $f_k(p)$ has a power-law fall for large $p$, i.e., $f_k(p) \sim p^{-(D+1)}$, known as Porod's law~\cite{porod1982small,op88}.

We probe the time correlations in the system via the autocorrelation function, defined as~\cite{PuriWad09,Bray94}
\be
\label{auto1}
A(t,t_{{\rm w}}) = \frac{1}{N} \sum_{i=1}^N \left[\overline{\left<s_{i}(t) s_{i}(t_{{\rm w}})\right>} - \overline{\left<s_{i}(t)\right>} ~ \overline{\left<s_{i}(t_{{\rm w}})\right>}\right],
\ee
where $t_w~(<t)$ is the {\it waiting time}, which describes the age of the system. In coarsening systems, for $t-t_w \gg t_w$, the quantity $A(t,t_w)$ usually has the following dynamical scaling form:
\be
\label{auto3}
A(t,t_{{\rm w}}) = g\left(\frac{R(t)}{R(t_{{\rm w}})}\right),
\ee
where $g(y)$ is the scaling function.

\section{Numerical Results}
\label{s3}

In this section, we discuss our simulation results for the $1D$ RFLRIM. We take a large system size $N = 2^{20}$ ($\sim 1$ million spins) to ensure that the results are free from finite-size effects. All statistical quantities obtained in this paper are averaged over 50-100 independent runs, each with different initial conditions and disorder configurations.

\subsection{Domain Growth Law}

Domain growth is characterized by the diffusion of interfaces in order to lower the net free energy of the system. In the presence of a quenched random field, these interfaces become locally pinned leading to the creation of free energy barriers. The subsequent interfacial motion is an activated process rather than free diffusion. In the context of the $1D$ RFIM with NN interactions where interfaces are point defects, Fisher et al.~\cite{PhysRevE.64.066107} proposed that an interface can be treated as a random walker in a random static force field of the Sinai type. Since there are multiple interfaces in a coarsening system, they simultaneously diffuse and annihilate upon meeting.

\begin{figure}[t!]
	\centering
	\includegraphics[width=0.65\linewidth]{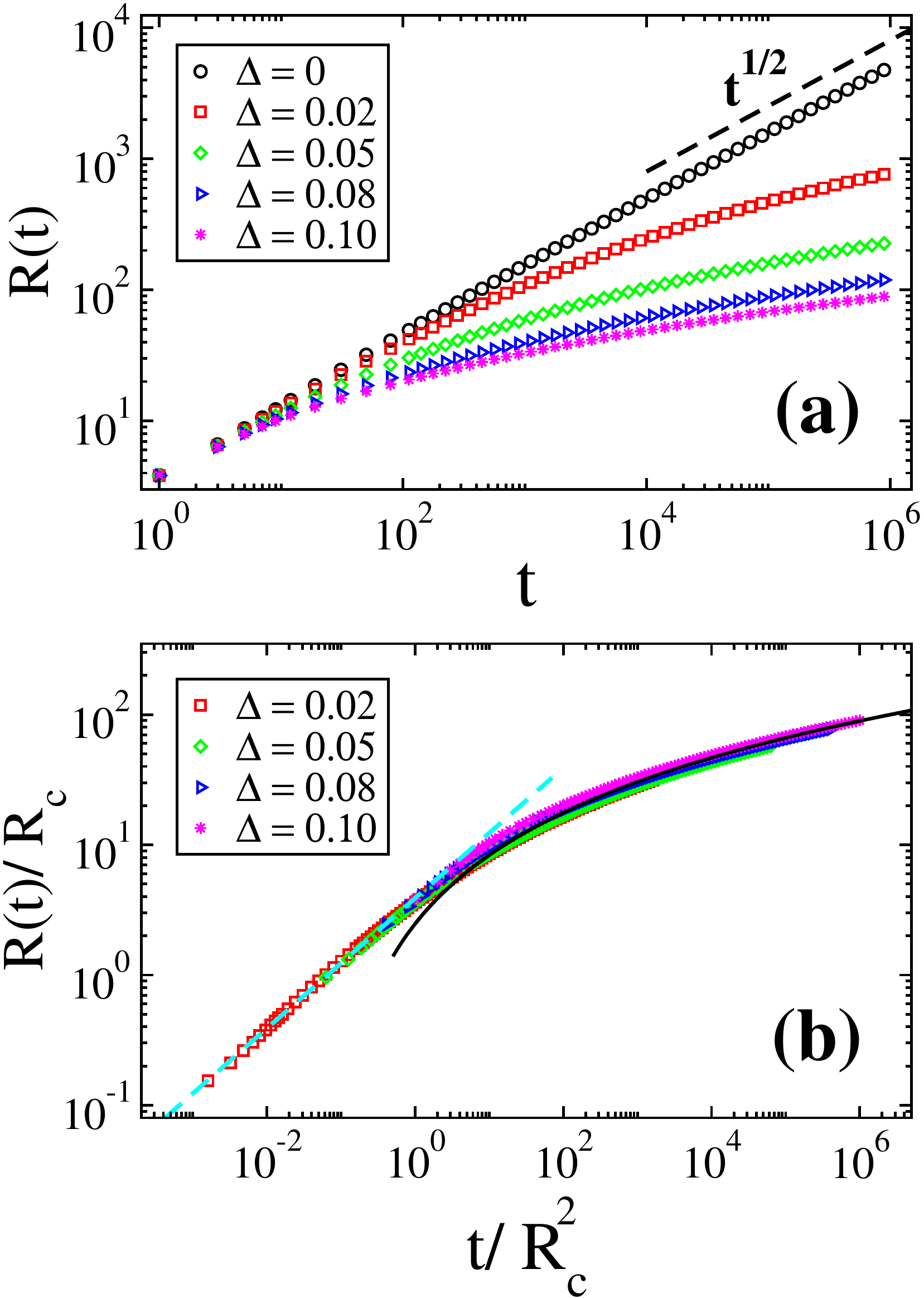}
	\caption{(a) Plot of characteristic length scale $R(t)$ vs. time $t$ on a log-log scale, for a quench to $T = 0.1$ of the $1D$ NN RFIM model. The data sets correspond to different disorder amplitudes $\Delta$. The dashed line denotes $R(t) \sim t^{1/2}$. (b) Plot of $R(t)/R_{\rm c}$ vs. $t/R_{\rm c}^2$ for the data in (a), where $R_{\rm c}=\left( T/\Delta \right)^2$ is the crossover scale (see text). The dashed line is a fit to the pre-asymptotic regime with $\mathcal{L}(x) \sim 3.9 \times x^{1/2}$. The solid line denotes a fit to the asymptotic regime with $\mathcal{L}(x) \sim 0.32 \times \left[ \ln (15.8 x) \right]^2$.}
	\label{nn}
\end{figure}

The mean-squared displacement of a random walk in the Sinai potential is given by $\langle x^2(t) \rangle \simeq \left( \ln t \right)^4$~\cite{doi:10.1137/1127028,BOUCHAUD1990285}. This can be understood as follows. For a displacement $x$, the walker faces a barrier of typical height $\Delta \sqrt x$, which requires an Arrhenius-type escape time $t\sim \exp ( \Delta \sqrt x /T)$. Solving for $x$, one obtains the above result. The growth law in the RFIM follows this Sinai diffusion regime at asymptotic times: $R(t) \sim \left( \ln t \right)^2$~\cite{PhysRevE.65.046114}. In the early regime, thermal energy dominates the free energy barriers and the interfacial motion occurs via free diffusion as in the pure Ising model. The characteristic crossover length can be obtained by comparing the thermal energy with the typical barrier height~\cite{corberi2011growth,PhysRevE.65.046114}, $\Delta \sqrt{R_c} \simeq T$, leading to
\be
\label{barr}
R_{\rm c} = \left( \frac{T}{\Delta} \right)^2 .
\ee
This crossover behavior can be captured in the scaling form~\cite{PhysRevE.65.046114}
\be
\label{interp}
R(t,R_{\rm c}) = R_{\rm c} \mathcal{L}\left( \frac{t}{R_{\rm c}^2} \right),
\ee
where the scaling function $\mathcal{L}(v)$ is
\bea
\mathcal{L}(v) &\sim& v^{1/2}, \quad \quad v \ll 1, \nonumber \\
&\sim& ( \ln ~v )^2, \quad v \gg 1 .
\eea

\begin{figure}[t!]
	\centering
	\includegraphics[width=0.85\linewidth]{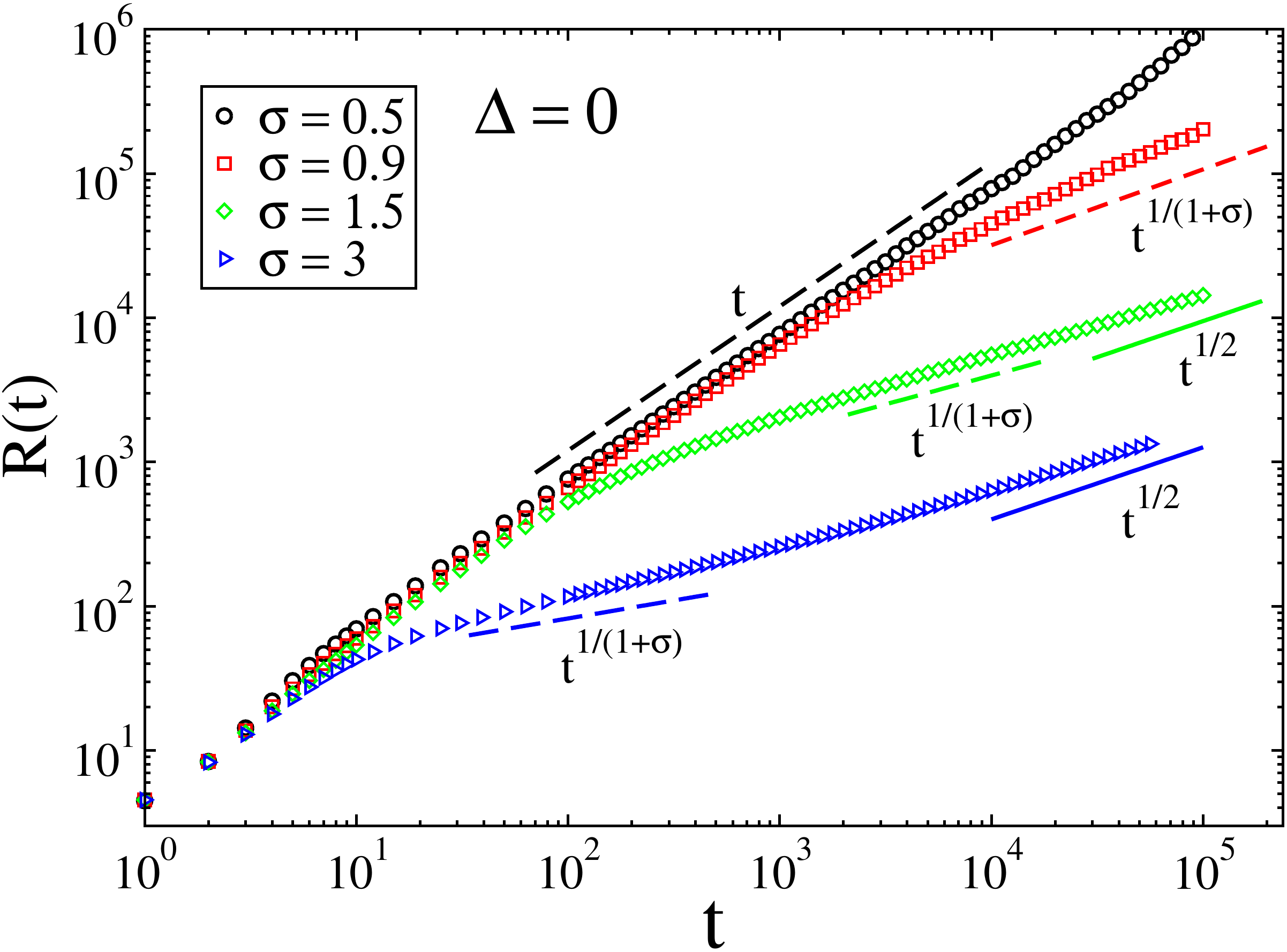}
	\caption{Plot of $R(t)$ vs. $t$ on a log-log scale for pure LR systems with various $\sigma$-values. We quench the system to $T = 10^{-3}$. The dashed and solid lines represent the expected pre-asymptotic and asymptotic growth laws (see text).}
	\label{lrim}
\end{figure}

Let us first benchmark the above result for the RFIM with NN interactions. In Fig.~\ref{nn}(a), we plot $R(t)$ vs. $t$ (on a log-log scale) for a quench to $T = 0.1$ of the NN model. When disorder $\Delta$ is zero, i.e., pure case, the growth law exhibits a power law, $R(t) \sim t^{1/2}$. For non-zero disorder values, diffusive growth is observed only at early times. At late times, the growth slows down. In the late stage of the simulations, the growth becomes much slower, indicated by a downward bending on the log-log scale of Fig.~\ref{nn}(a). To investigate the crossover quantitatively, we plot in Fig.~\ref{nn}(b) the scaling variables $R(t)/R_{\rm c}$ vs. $t/R_{\rm c}^2$ for the data in Fig.~\ref{nn}(a). For the validity of Eq.~\eqref{interp}, a data collapse should occur in such a plot, which is indeed observed in Fig.~\ref{nn}(b). The data in limiting regimes shows excellent agreement with the expected forms (see the caption for details).

What happens to the above scenario if the spins in the system have LR interactions? In the following, we answer this question using extensive numerical simulations. However, before doing that, it is useful to briefly discuss the behavior of the pure system with LR interactions. The extended interactions exert a drift force on the moving interfaces in the system, which may change the growth law with respect to the NN case. Bray and Rutenberg~\cite{PhysRevE.49.R27,PhysRevE.50.1900} derived the modified growth law using a continuum approach based on a Ginzburg-Landau free energy. (The same result was derived for the Ising model by Corberi et al.~\cite{corberi2019one}.) According to their predictions, $R(t)$ for a non-zero temperature quench is given as 
\bea
R(t) &\sim& t^{1/(1+\sigma)}, \quad \sigma < 1 , \nonumber \\
&\sim& t^{1/2}, \quad \sigma > 1 .
\eea
Corberi et al.~\cite{corberi2019one} also found that this asymptotic regime is preceded by several pre-asymptotic regimes. For $\sigma < 1$, there is a ballistic growth regime, $R(t) \sim t$, at early time scales. For $\sigma > 1$, there is a slow regime sandwiched between the ballistic and the asymptotic regimes, where $R(t) \sim t^{1/(1+\sigma)}$. The crossover lengths were also determined. The ballistic regime ends when 
\be
R(t) \simeq R_b \sim \left (\frac{2}{\sigma T}\right) ^{1/\sigma} .
\ee
For $\sigma >1$, the asymptotic regime [$R(t) \sim t^{1/2}$] starts when
\be
R(t) \simeq R_d \sim \left (\frac{2}{\sigma T}\right) ^{1/(\sigma-1)} .
\ee
Notice that the duration of all the pre-asymptotic regimes decreases upon increasing $T$ or $\sigma$. In Fig.~\ref{lrim}, we benchmark the above growth law for quench temperature $T = 10^{-3}$. We choose such a low temperature so as to clearly delineate the pre-asymptotic regimes.

\begin{figure}[t!]
	\centering
	\includegraphics[width=0.95\linewidth]{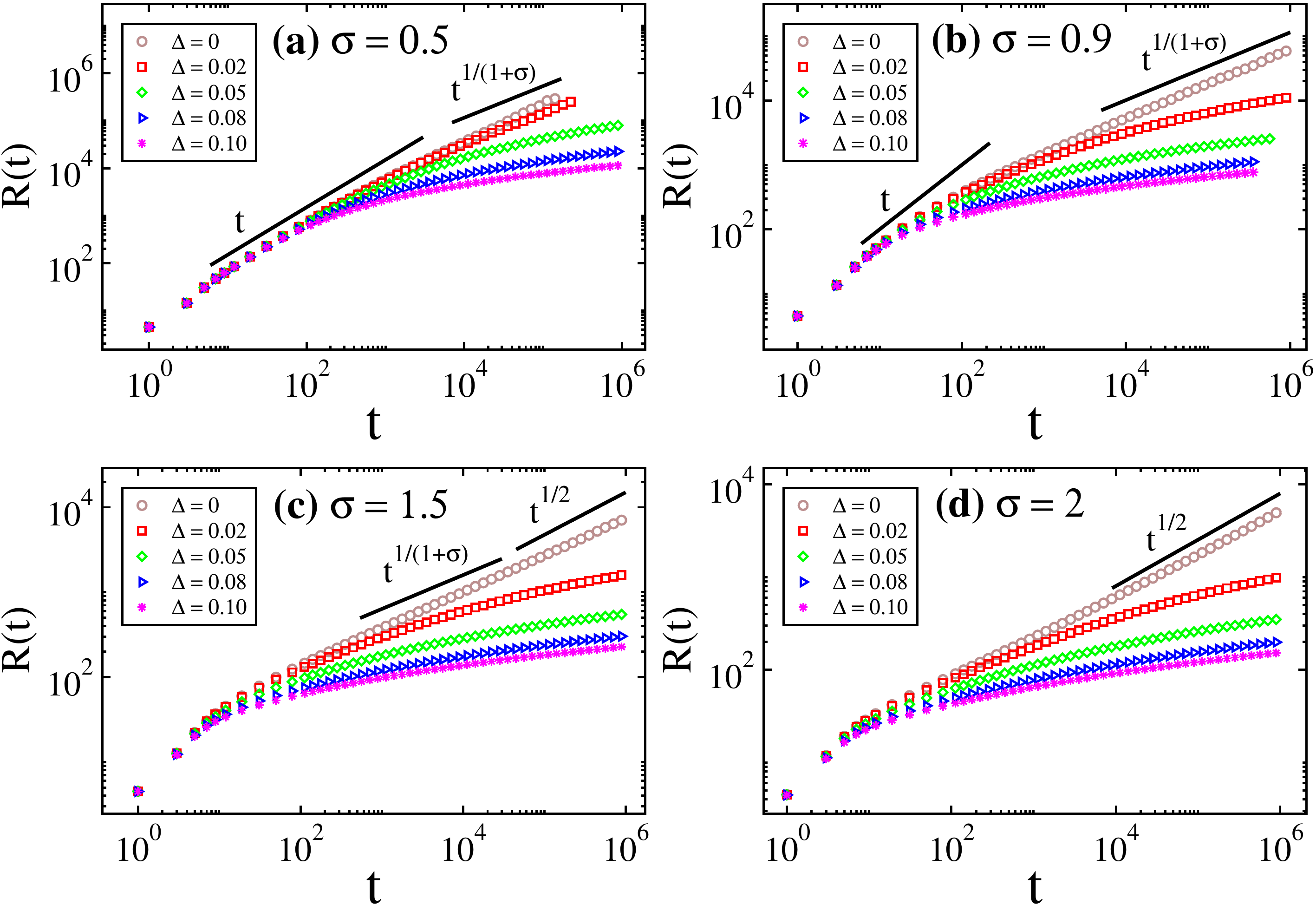}
	\caption{Plot of $R(t)$ vs. $t$ for a quench to $T = 0.1$ of the RFLRIM with various $\Delta$, and $\sigma =$ (a) 0.5, (b) 0.9, (c) 1.5, (d) 2. The solid lines in different frames represent the expected growth laws in the pure case.}
	\label{rflr}
\end{figure}

\begin{figure}[t!]
	\centering
	\includegraphics[width=0.95\linewidth]{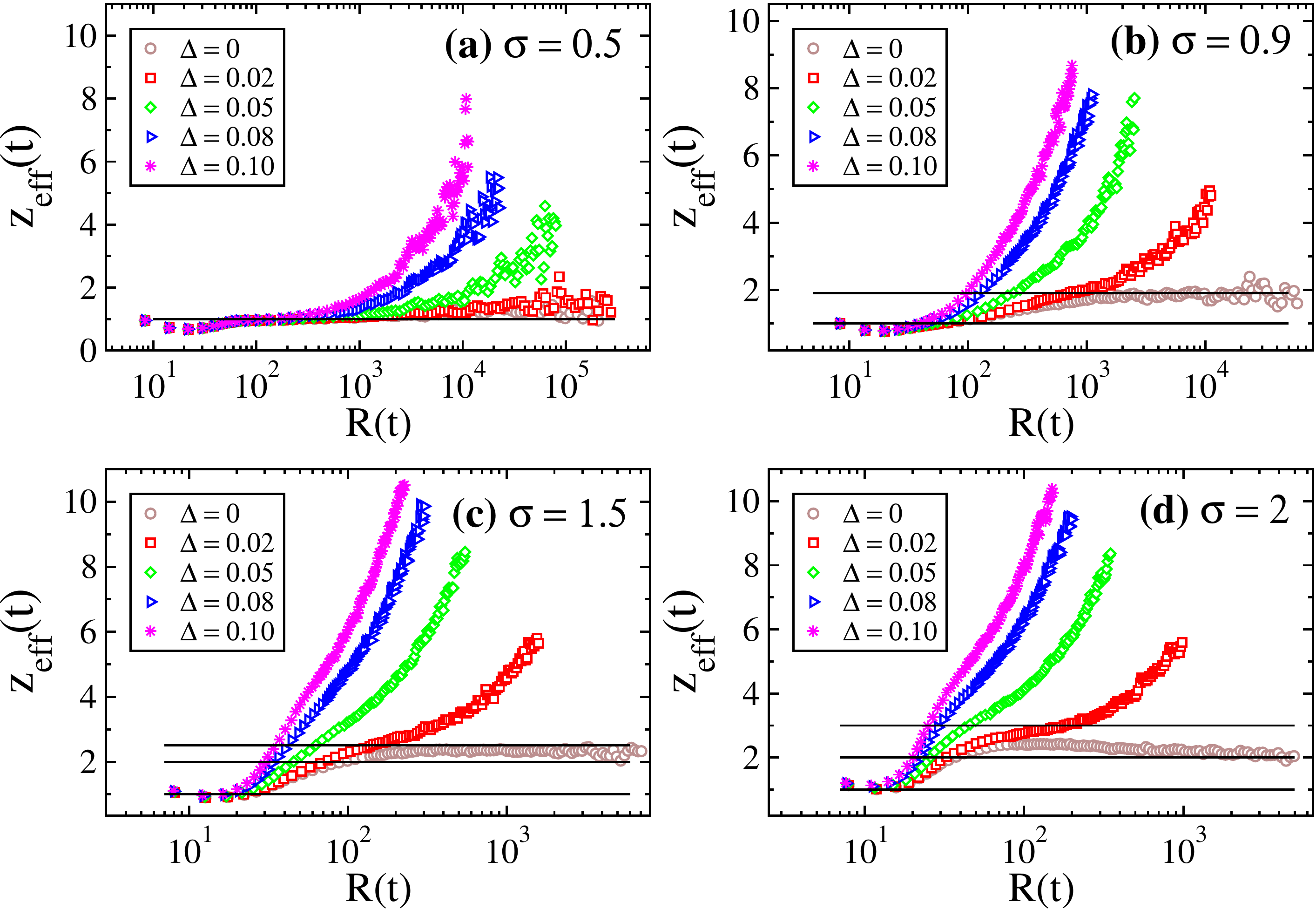}
	\caption{Plot of the effective growth exponent $z_{\rm eff}$ vs. $R(t)$ for the data sets in Fig.~\ref{rflr}, on a log-linear scale. The horizontal lines represent the pre-asymptotic and asymptotic exponents for the pure case.}
	\label{rflrexp}
\end{figure}

After recalling the behavior of the pure system, we move to our core results, i.e., the ordering kinetics of the $1D$ RFLRIM. In Fig.~\ref{rflr}, we plot $R(t)$ vs. $t$ for a quench to temperature $T = 0.1$ of LR systems with various values of $\sigma$ [see panels (a)--(d)] and $\Delta$. We have chosen 2 values of $\sigma > 1$ and 2 values of $\sigma < 1$, as $\sigma = 1$ is a significant boundary for the pure case. (The corresponding plots of effective exponent $z_{\rm eff}$ vs. $R(t)$ are shown in Fig.~\ref{rflrexp}.) Let us start our discussion from Fig.~\ref{rflr}(b), corresponding to $\sigma = 0.9$. For $\Delta = 0$, the growth is ballistic at early times, with a crossover to $R(t) \sim t^{1/(1+\sigma)}$ at late times (indicated by power laws with different slopes in the same figure). For small non-zero $\Delta$, the domains initially grow as in the pure case, i.e., the growth law is $R(t) \sim t$ with a crossover to $R(t) \sim t^{1/(1+\sigma)}$. We see a further crossover to much slower growth at late times. With further increase in $\Delta$, the power-law regimes shrink to smaller time-windows, and the crossover to asymptotic slow growth occurs much earlier. As discussed above, this slow growth is a result of the quenched random field. In Fig.~\ref{rflrexp}(b), the quantity $z_{\rm eff}$ shows an upward trend after an initial flat behavior, which is interpreted~\cite{corberi2011growth,corberi2012crossover} as a signature of logarithmic growth. We remark that the same scenario holds for other values of $\sigma$, as shown in Figs.~\ref{rflr}(a),(c) and (d) and Figs.~\ref{rflrexp}(a),(c) and (d). The only difference is that the various regimes can widen, shrink or even disappear as $\sigma $ is varied.

\begin{figure}[t!]
	\centering
	\includegraphics[width=0.85\linewidth]{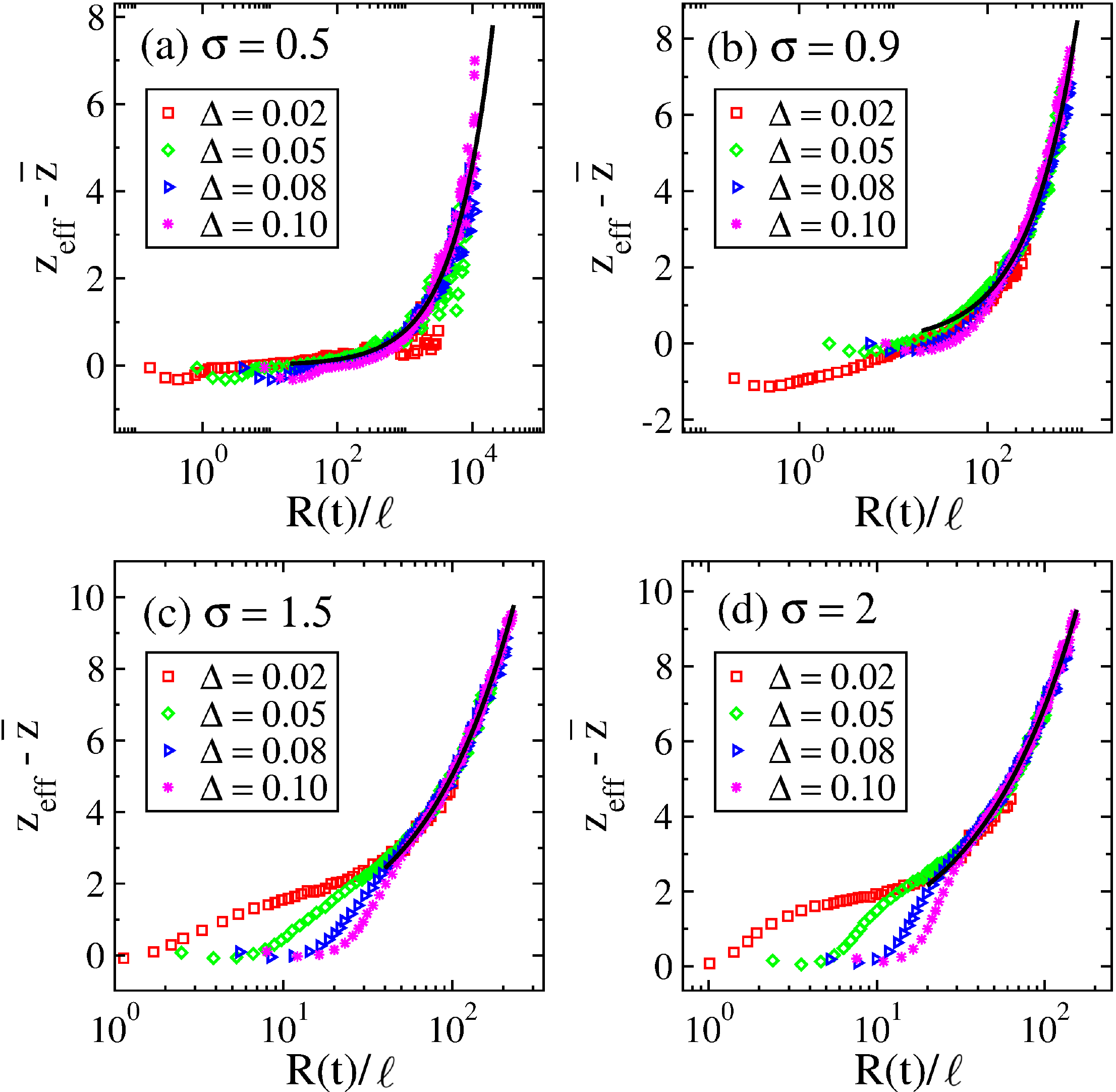}
	\caption{Scaling plot of $z_{\rm eff} - \bar{z}$ vs. $R(t)/\ell$ for the data in Fig.~\ref{rflrexp}. The panels correspond to $\sigma =$ (a) 0.5, (b) 0.9, (c) 1.5, (d) 2. The crossover length $\ell(\Delta)$ is chosen to enable a data collapse. The parameter $\bar{z}$ is the disorder-independent pre-asymptotic growth: $R(t) \sim t^{1/\bar{z}}$. The solid line in different panels is the best power-law fit: $z_{\rm eff} - \bar{z} \propto (R/\ell)^{\psi}$.}
	\label{rflr2}
\end{figure}

We now present a more quantitative analysis of the asymptotic growth. Considering the behavior of the NN model and the Huse-Henley argument \cite{corberi2011growth,PhysRevLett.54.2708}, we make the working hypothesis that the asymptotic growth is logarithmic:
\be
R(t)\sim \left [ \ln(t/\tau)\right ]^{\alpha(\sigma)} .
\label{workhyp}
\ee
Here, $\tau$ sets the time-scale, and we know that $\alpha (\sigma=\infty)=2$ for the NN model~\cite{PhysRevE.64.066107}. Corberi et al. \cite{corberi2011growth,corberi2012crossover} have developed a novel scaling method to study the crossover from the pre-asymptotic power-law regime to the asymptotic logarithmic regime. This has been detailed in several publications from our group, and we do not replicate the arguments here (see Ref.~\cite{PhysRevE.104.044123} for example). Their method is based on a scaling plot of $z_{\rm eff} - \bar{z}$ vs. $R(t)/\ell (\Delta)$, where $\bar{z}$ is the pre-asymptotic power-law exponent, and $\ell (\Delta)$ is chosen to facilitate a data collapse (see Table~\ref{t1}). In Fig.~\ref{rflr2}, we undertake such a plot of the data sets in Fig.~\ref{rflrexp}. We see a reasonable data collapse for each value of $\sigma$. The corresponding power-law fit to the scaled data [i.e., $z_{\rm eff} - \bar{z} = a(R/\ell)^{\psi (\sigma)}$], shown as a solid line in Fig.~\ref{rflr2}(a)-(d), implies a logarithmic growth of the domain size:
\be
R(t) \sim \left [ \ln(t/\tau)\right ]^{1/\psi(\sigma)} , \quad \tau = \ell^{\bar{z}} .
\ee
Thus, we identify the logarithmic exponent as $\alpha(\sigma) = 1/\psi(\sigma)$. The resultant values of $\ell (\Delta)$ and $\alpha(\sigma)$ are presented in Tables~\ref{t1} and \ref{t2} respectively.

Our estimates of $\alpha(\sigma)$ are not very accurate, and we do not have data for enough $\sigma$-values, to enable us to make a precise statement about the $\sigma$-dependence of $\alpha$. However, it is clear that $\alpha(\sigma)$ is smaller than the value $\alpha(\sigma=\infty) = 2$ of the NN case. This is surprising and requires detailed clarification, which we provide below.

\begin{figure}[t]
	\centering
	\includegraphics[width=0.7\linewidth]{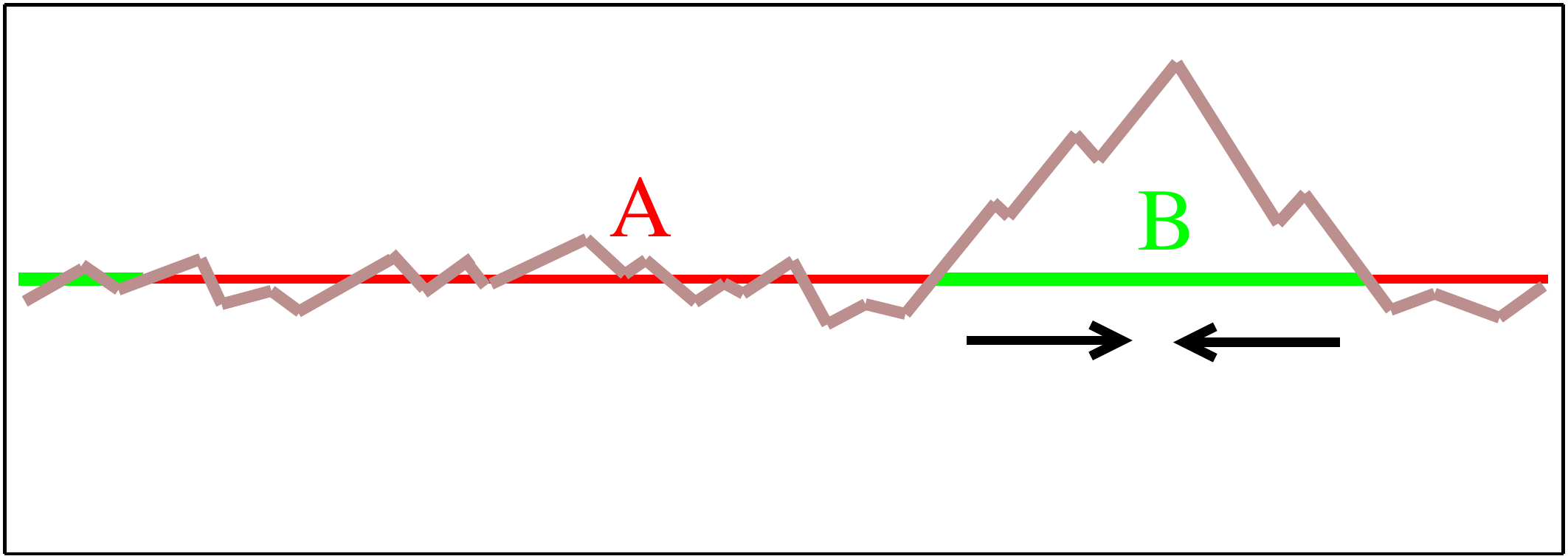}
	\caption{Pictorial representation of the $1D$ RFLRIM. The red (thin) and green (thick) lines correspond to domains of up and down spins, respectively. The energy height barrier profile is drawn in brown. The arrows represent the force exerted by the LR interaction, which promotes the closure of the small green domain in the pure case.}
	\label{fig_barrier}
\end{figure}

In the pure case, the LR interaction speeds up (or, at most, leaves unchanged) the growth law. This is due to an additional force caused by far away spins that promotes the closure of smaller domains~\cite{corberi2019one}. However, with the quenched random field, small domains, like the one denoted by {\bf B} in Fig.~\ref{fig_barrier}, can be eliminated only if the random field is not too unfavorable to the process. Hence, after some time, the small domains that survive are those which the random field itself makes harder to close. In a situation like the one in Fig.~\ref{fig_barrier}, the force associated with the LR interactions suppresses the most efficient coarsening mechanism, thereby slowing down growth. This is a possible explanation for the smaller values of $\alpha(\sigma)$ seen in Table~\ref{t2}.

\begin{table}[t!]
	\begin{center}
		\begin{tabular}{| c | c | c | c |}
			\hline 
			& $\sigma = 0.5$ & $\sigma = 0.9$ & $\sigma = 1.5$ \\
			\hline
			$\Delta=0.02$ & 50.0 & 41.0 & 15.42 \\
			\hline
			$\Delta=0.05$ & 10.0 & 4.0 & 3.25 \\
			\hline
			$\Delta=0.08$ & 2.1 & 1.5 & 1.46 \\
			\hline
			$\Delta=0.1$ & 1.0 & 1.0 & 1.0 \\
			\hline
		\end{tabular}
	\end{center}
	\caption{Scaling length $\ell(\Delta)$ for different $\sigma$ in the $D=1$ RFLRIM. These values are chosen so as to facilitate a data collapse in Fig.~\ref{rflr2}).}
	\label{t1}
\end{table}

\begin{table}[t!]
	\begin{center}
		\begin{tabular}{| c | c |}
			\hline 
			$\sigma$ & $\alpha$ \\
			\hline
			0.5 & 1.31 (5) \\
			\hline
			0.9 & 1.17 (1) \\
			\hline
			1.5 & 1.26 (1) \\
			\hline
			2.0 & 1.38 (2) \\
			\hline
			$\infty$ & 2.0 \\
			\hline
		\end{tabular}
	\end{center}
	\caption{Logarithmic growth exponent $\alpha$ for different $\sigma$ in the $D=1$ RFLRIM. These values are obtained from the scaling plots of $z_{\rm eff} - \bar{z}$ vs. $R(t)/\ell$ (see Fig.~\ref{rflr2}), as described in the text.}
	\label{t2}
\end{table}

\subsection{Dynamical Scaling and Universality of $C(r,t)$}

\begin{figure}[t!]
	\centering
	\includegraphics[width=0.75\linewidth]{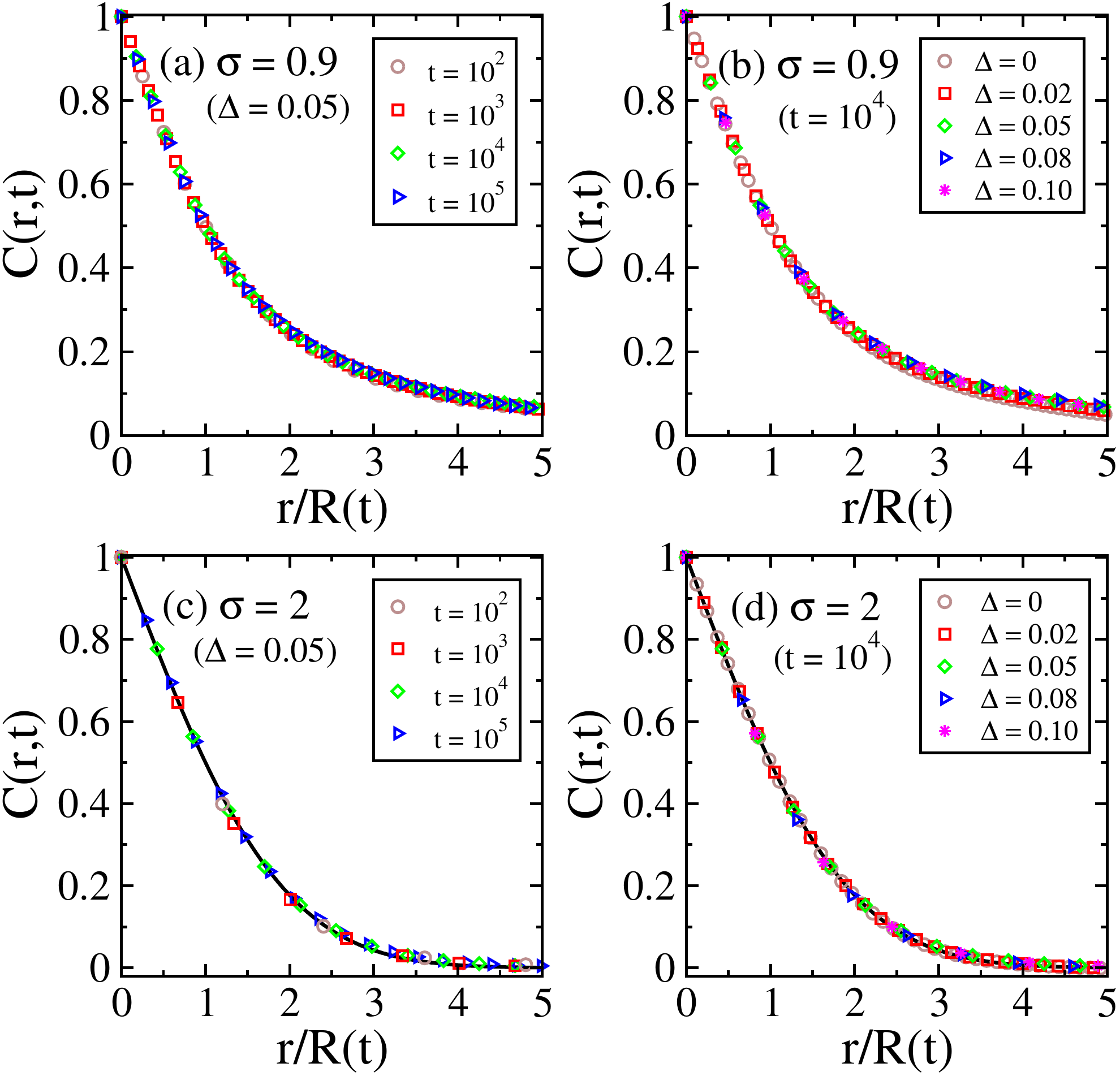}
	\caption{Scaling plots of the equal-time correlation function, $C(r,t)$ vs. $r/R(t)$, for a quench to $T = 0.1$ of the RFLRIM with (a)-(b) $\sigma = 0.9$, and (c)-(d) $\sigma = 2$. The frames (a) and (c) show data sets at different $t$ for a system with $\Delta = 0.05$. The frames (b) and (d) show data sets at fixed $t = 10^4$ with various $\Delta$. The solid curve in (c)-(d) denotes the scaling function for the pure NN model, i.e., Eq.~\eqref{glaub} with $c \simeq 0.55$.}
	\label{dscale}
\end{figure}

\begin{figure}[t!]
	\centering
	\includegraphics[width=0.5\linewidth]{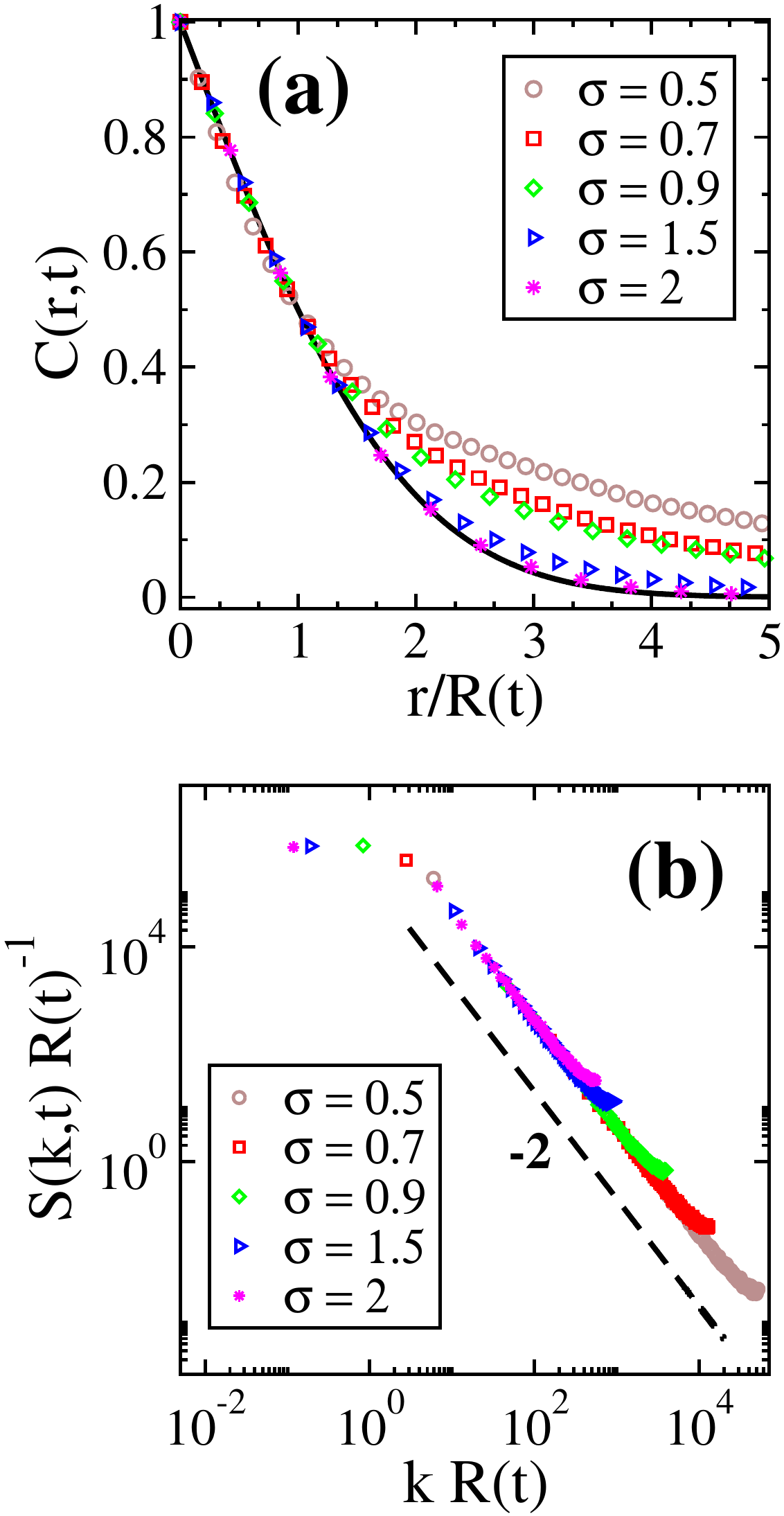}
	\caption{Scaling plots of (a) $C(r,t)$ vs. $r/R(t)$; and (b) $S(k,t) R(t)^{-1}$ vs. $k R(t)$ for a quench of the RFLRIM to $T = 0.1$. We plot data for various $\sigma$-values with fixed $t = 10^4$ and $\Delta = 0.05$. The solid line in (a) denotes Eq.~\eqref{glaub}. The dashed line in (b) denotes the Porod law: $S(k,t) \sim k^{-2}$.}
	\label{dscale2}
\end{figure}

Another important aspect of domain growth is dynamical scaling, which we explore in this section using the spatial correlation function $C(r,t)$, defined in Eq.~(\ref{eq5}). In Fig.~\ref{dscale}(a), we plot $C(r,t)$ vs. $r/R(t)$ for distinct simulation times (across 3 decades) for the RFLRIM with $\sigma = 0.9$ and $\Delta = 0.05$. The data sets exhibit excellent data collapse on a single scaling function, demonstrating the validity of dynamical scaling [Eq.~(\ref{cf_scale})] in the system. Fig.~\ref{dscale}(b) testifies the SU nature of this scaling function for $\sigma = 0.9$. Here, we present data for $C(r,t)$ vs. $r/R(t)$ for different $\Delta$ at a fixed $t=10^4$ MCS. All the data sets collapse onto the same scaling function, showing the validity of SU. We show analogous plots for $\sigma = 2$ in Figs.~\ref{dscale}(c)-(d). Again, the data sets fall on a single scaling function in both frames. The solid lines in Figs.~\ref{dscale}(c)-(d) denote the scaling function for the $1D$ pure NN model with Glauber dynamics~\cite{Bray94}:
\be
\label{glaub}
f(x) = {\rm erfc} \left( c x \right),
\ee
where the constant $c \simeq 0.5$. The data in Figs.~\ref{dscale}(c)-(d) agree very well with this function, i.e., for large $\sigma~(>1)$ the scaling function falls in the NN universality class. (This is not the case for $\sigma=0.9$.) 
To complete our study of $C(r,t)$, we superpose scaled data for different values of $\sigma$ in Fig.~\ref{dscale2}(a). The data sets correspond to fixed $\Delta = 0.05$ and $t = 10^4$ MCS. For $r < R(t)$, the data for different $\sigma$ and the NN case match well, showing that the geometry of domains at small scales is unaffected by the extended interactions. However, at large scales ($r > R(t)$), the decay of $C(r,t)$ has a strong dependence on $\sigma$. For $\sigma < 1$, the fall is relatively slow and depends on the value of $\sigma$. For $\sigma > 1$, the scaling function is consistent with that for the NN case. We conjecture that (as in the pure case) the scaling function in the disordered case varies continuously with $\sigma$ for $\sigma < 1$. For $\sigma > 1$, the function is independent of $\sigma$, and is equivalent to the pure NN scaling function.

We also test the dynamical scaling of the structure factor (not shown here). We find excellent scaling of the data, and the structure-factor tails show the $1D$ Porod law: $S(k,t) \sim k^{-2}$ for large $k$ (for all $\sigma$ and $\Delta$-values presented in this paper). This is seen in Fig.~\ref{dscale2}(b), where we plot the scaled structure factors [$S(k,t) R(t)^{-1}$ vs. $k R(t)$] corresponding to the data sets shown in Fig.~\ref{dscale2}(a). The Porod law is characteristic of scattering from sharp interfaces in the system. In higher dimensions, it is known that disordered spin models exhibit a generalization of the Porod law or {\it fractal Porod law} with non-integer exponents \cite{skb11,skb14}. This is a consequence of the roughening of the interfaces into fractals. In the $1D$ case considered here, we do not see a fractal Porod law as the interfaces are point defects.

\subsection{Dynamical Scaling and Universality of $A(t,t_w)$}

\begin{figure}[t!]
	\centering
	\includegraphics[width=0.72\linewidth]{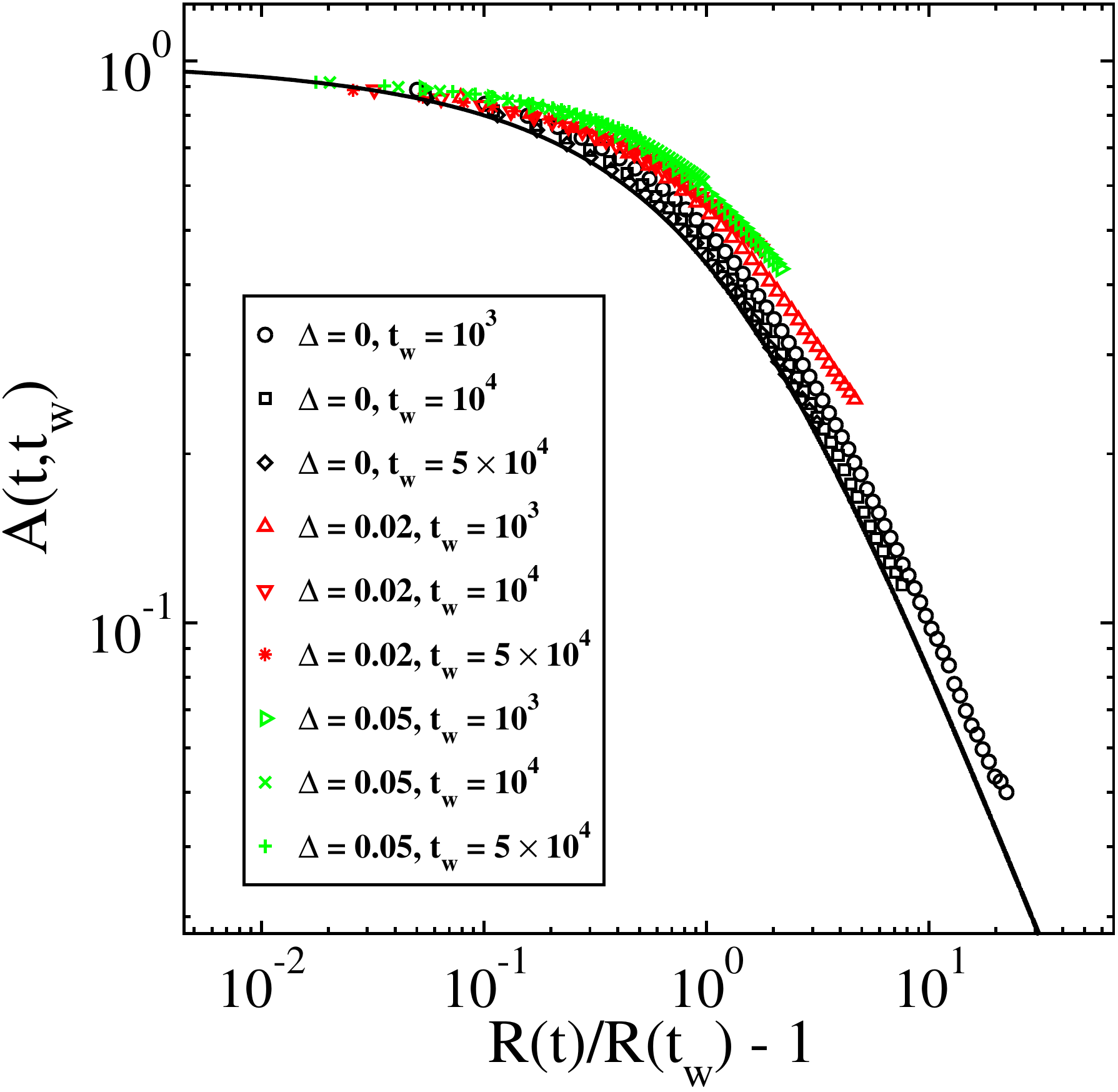}
	\caption{Plot of autocorrelation function, $A(t,t_{\rm w})$ vs. $R(t)/R(t_w) - 1$, for the RFLRIM. We quench a system with $\sigma = 2$ to $T=0.1$, and show data for various $\Delta$ and $t_w$. The solid curve denotes the scaling function in Eq.~\eqref{nnauto} for the pure NN model.}
	\label{autoLarge}
\end{figure}

\begin{figure}[t!]
	\centering
	\includegraphics[width=0.8\linewidth]{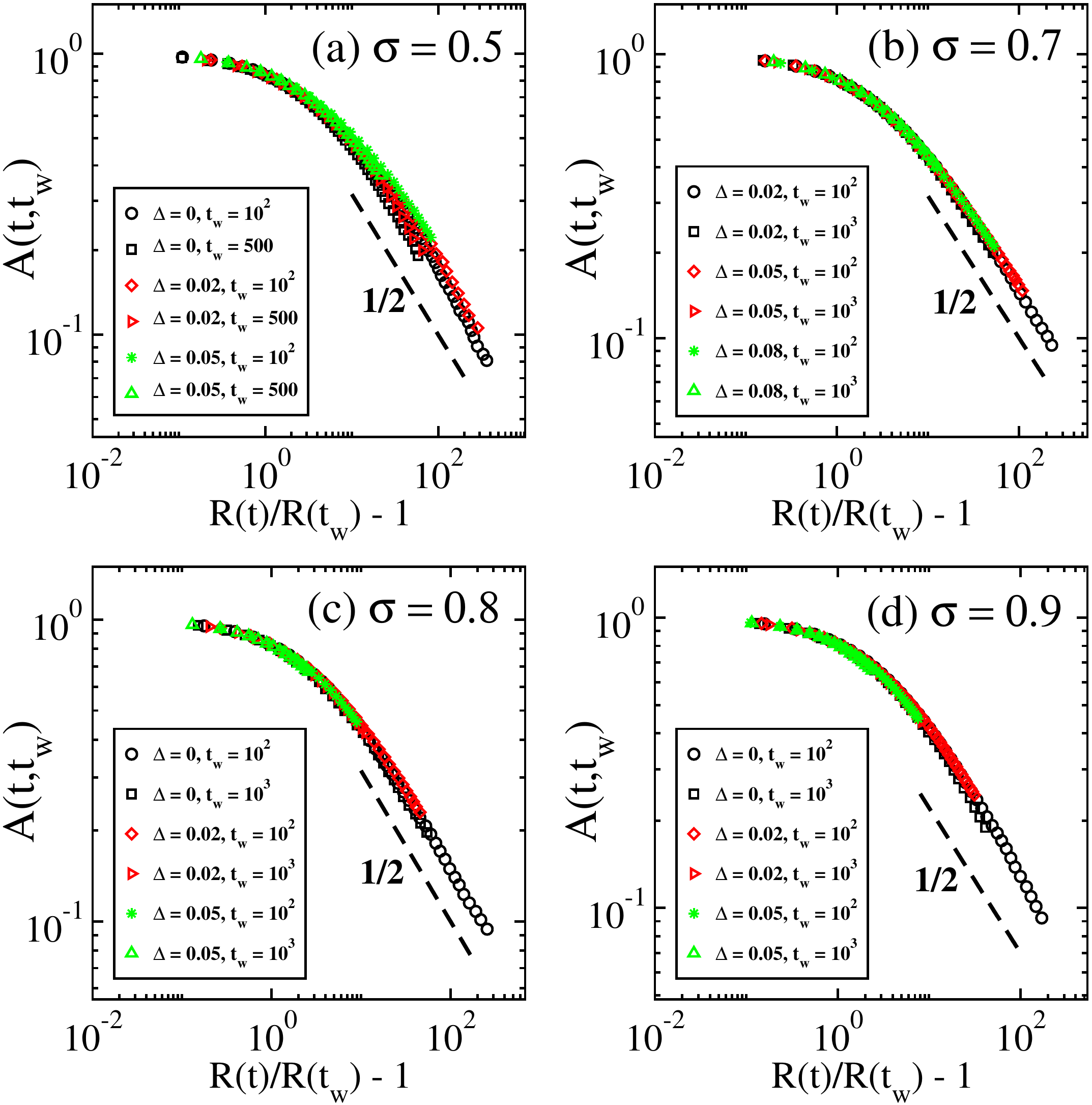}
	\caption{Plot of $A(t,t_{{\rm w}})$ vs. $R(t)/R(t_w) - 1$ for the RFLRIM with $\sigma =$ (a) 0.5, (b) 0.7, (c) 0.8, (d) 0.9. We present data for various $\Delta$ and $t_w$. The system with $\sigma = 0.7$ is quenched to $T = 0.5$, whereas the other frames correspond to $T = 0.1$. The dashed line denotes $g(y) \sim y^{-\lambda}$ with $\lambda = 1/2$.}
	\label{auto}
\end{figure}

The two-time quantities offer a testing ground for SU in the time correlations. As mentioned earlier, there are several known examples of SU violation in two-time quantities \cite{PhysRevE.65.046114,corberi2012crossover,PhysRevE.104.044123}. Here, we specifically investigate the autocorrelation function $A(t,t_w)$, defined in Eq.~(\ref{auto1}). Before showing our results, it is useful to summarize what is known in the $1D$ NN case. In the pure system, $A(t,t_w)$ obeys dynamical scaling [Eq.~(\ref{auto3})] with a scaling function~\cite{Bray94}:
\be
\label{nnauto}
g\left( y \right) = \frac{2}{\pi} ~{\rm arcsin} \left(\sqrt{\frac{2}{1+y^2}} \right).
\ee
This function decays as $y^{-\lambda}$ for large $y$, where $\lambda$ defines the so-called Fisher-Huse exponent \cite{fh88}. In this case, it turns out to be $\lambda=1$. The same scaling function was shown to describe $A(t,t_w)$ when LR interactions are added with $\sigma>1$~\cite{Corberi_2019}. The addition of the random field in the NN case modifies $g(y)$ to~\cite{PhysRevE.64.066107,PhysRevE.65.046114}
\be
\label{nnautoh}
g\left( y \right) = \frac{4}{3\sqrt y}-\frac{1}{3y} , \quad y \gg 1 ,
\ee
which differs from $g(y)$ of Eq.~(\ref{nnauto}), indicating a violation of SU. In particular, note that the asymptotic decay of $g(y)$ in Eq.~(\ref{nnautoh}) yields a Fisher-Huse exponent $\lambda = 1/2$, distinct from the $\lambda = 1$ observed in Eq.~(\ref{nnauto}). It was shown in~\cite{PhysRevE.65.046114} that the form in Eq.~(\ref{nnauto}) is observed pre-asymptotically also in the presence of the random field, while the true asymptotic form in Eq.~(\ref{nnautoh}) is approached only at late times by means of a slow, smooth crossover.

In our RFLRIM system, we find that the scenario described for the NN case is applicable for larger values of $\sigma$. This can be observed in Fig.~\ref{autoLarge}, where we plot $A(t,t_{{\rm w}})$ vs. $R(t)/R(t_w) - 1$ for $\sigma=2$, and various $\Delta$ and $t_w$. The data sets for $\Delta = 0$ approach the NN result with increase in $t_w$. The weak $t_w$-dependence of the scaled data can be attributed to scaling corrections, i.e., earlier values of $t_w$ are not fully in the asymptotic regime~\cite{Corberi_2019}. The data sets with same non-zero $\Delta$ collapse well on top of each other, confirming dynamical scaling. Further, the data for non-zero $\Delta$ shows substantial departure from the pure case, which is expected due to the above discussion on the NN case. However, the curves presented in Fig.~\ref{autoLarge} may not be representative of the {\it true} scaling function in the presence of disorder, due to the slow crossover discussed above for the NN case. A similar behavior is observed for $\sigma = 1.5$ (not shown here). We believe that the NN scenario of SU violation applies for the LR system with $\sigma > 1$ also.

We next investigate the SU property for small $\sigma$. Before presenting our results, it is useful to recap what is known in the pure case. We have already mentioned that the scaling function $g(y)$ takes the NN form for $\sigma >1$. For $\sigma < 1$, $g(y)$ changes to a different $\sigma$-independent function, characterized by $\lambda=1/2$~\cite{Corberi_2019}.

Let us now discuss the RFLRIM with $\sigma < 1$. In Fig.~\ref{auto}, we present scaling plots for $A(t,t_w)$ in systems with various $\sigma$, at different values of $\Delta$ and $t_w$. For the validity of SU, all the data sets belonging to the same $\sigma$, irrespective of the value of $\Delta$, should collapse on a single scaling form. This is indeed clearly observed in the figure, other than for $\sigma=0.5$ in Fig.~\ref{auto}(a). We remark that the slight departure for $\sigma = 0.5$ does not disprove the SU of $g(y)$. This is because, for $\sigma=0.5$, data are not in the asymptotic stage, but in the pre-asymptotic ballistic regime [see Fig.~\ref{rflr}(a)].

\begin{figure}[t!]
	\centering
	\includegraphics[width=0.72\linewidth]{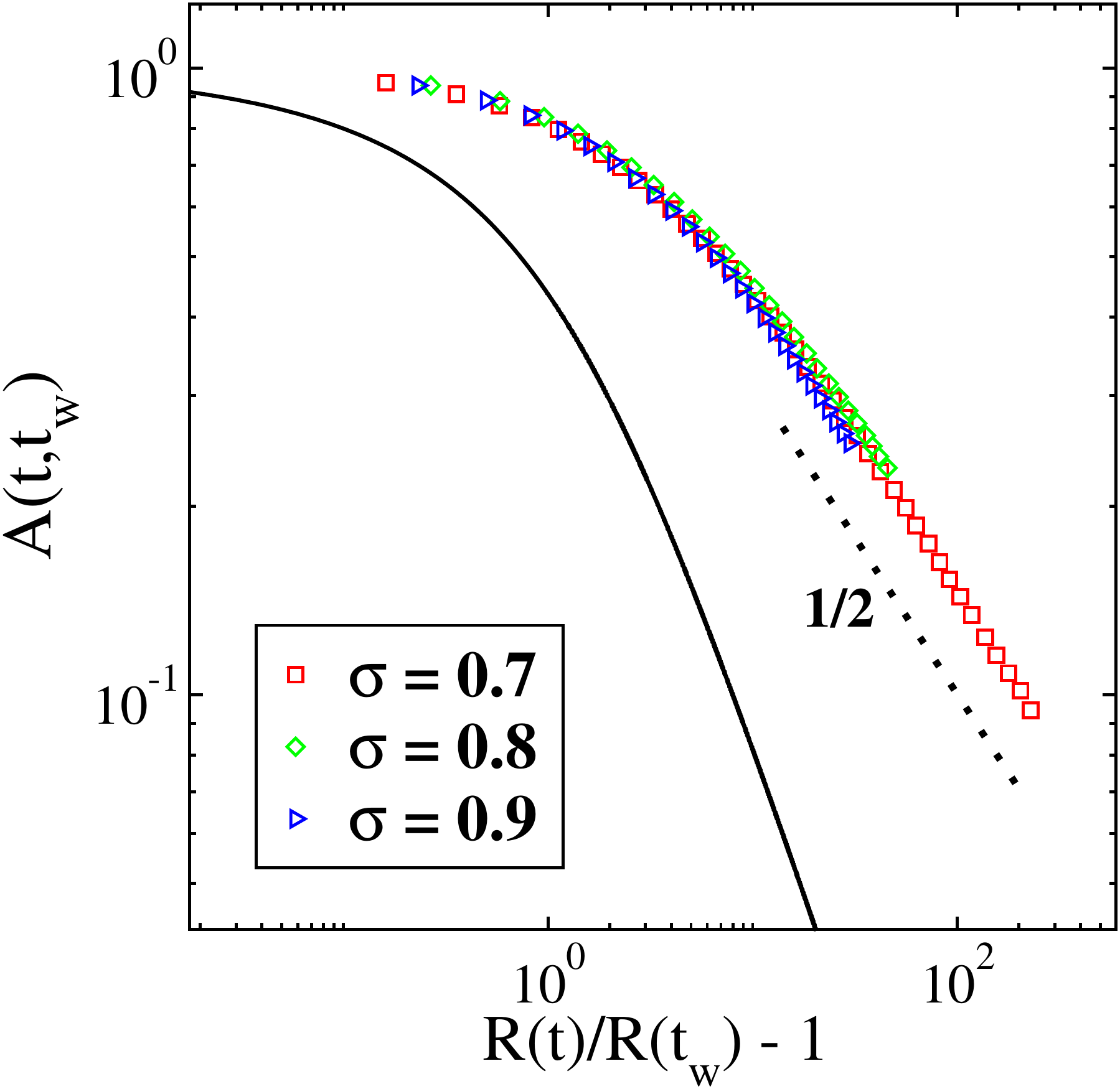}
	\caption{Plot of $A(t,t_{{\rm w}})$ vs. $R(t)/R(t_w) - 1$ for the RFLRIM with various $\sigma$. We set $\Delta = 0.02$ and $t_w = 100$. The system with $\sigma = 0.7$ is quenched to $T = 0.5$, while the quench temperature is $T = 0.1$ for $\sigma=0.8, 0.9$. The solid curve denotes Eq.~\eqref{nnauto}. The dotted line denotes $g(y) \sim y^{-\lambda}$ with $\lambda = 1/2$.}
	\label{autoSmall}
\end{figure}

In Fig.~\ref{autoSmall}, we plot $A(t,t_{{\rm w}})$ vs. $R(t)/R(t_w) - 1$ for various values of $\sigma < 1$ from Fig.~\ref{auto}. We consider a fixed $\Delta = 0.02$ and $t_w = 10^2$ MCS. The solid curve denotes the scaling function for the pure NN case. The scaling functions are $\sigma$-independent in this case. Further, the tail is consistent with $\lambda = 1/2$.

We conclude that the scaling function for $A(t,t_w)$ in the RFLRIM respects SU for $\sigma < 1$ with a $\sigma$-independent scaling function. On the other hand, SU is violated for $\sigma > 1$. For the convenience of the reader, we present an overview of our results for dynamical scaling of $C(r,t)$ and $A(t,t_w)$ in Table~\ref{t3}.

\begin{table}[t!]
	\begin{center}
		\begin{tabular}{| c | c | c | c |}
			\hline 
			{\it Quantity} & $\sigma$ & {\it $\Delta$-Independence (SU)} & {\it $\sigma$-Independence} \\
			\hline
			\multirow{2}{*}{$C(r,t)$} & $\sigma >1$ & \text{Yes} & \text{Yes} \\
			\cline{2-4}
			& $\sigma <1$ & \text{Yes} & \text{No} \\
			\hline
			\multirow{2}{*}{$A(t,t_w)$} & $\sigma >1$ & \text{No} & \text{No} \\
			\cline{2-4}
			& $\sigma <1$ & \text{Yes} & \text{Yes} \\
			\hline
		\end{tabular}
	\end{center}
	\caption{Properties of scaling functions for $C(r,t)$ and $A(t,t_w)$ in different ranges of $\sigma$.}
	\label{t3}
\end{table}

\section{Summary and Discussion}
\label{s4}

Let us conclude this paper with a summary and discussion of our results. We have investigated numerically the post-quench dynamics of the $1D$ random field Ising model (RFIM) with long-range (LR) interactions. To the best of our knowledge, this is the first study of ordering kinetics in a disordered system with LR interactions. Previous works of Fisher et al.~\cite{PhysRevE.64.066107} and Corberi et al.~\cite{PhysRevE.65.046114} showed that the interfaces in the RFIM with NN interactions diffuse in a quenched random potential of the Sinai type~\cite{doi:10.1137/1127028,BOUCHAUD1990285}, which yields an asymptotic growth law $R(t) \sim (\ln t)^2$. The extended interactions considered here exert an additional drift force on the interfaces moving in the Sinai environment, which changes the growth law.

Our data does not enable us to make a definitive statement about the functional form of the growth law. This is because the sluggish growth does not yield a significant increase in the length scale in the time-window of our simulations. However, we provide compelling evidence that a logarithmic law $R(t)\sim (\ln t)^{\alpha(\sigma)}$ is consistent with the data in the asymptotic time domain accessed by the simulations. Interestingly, the exponent $\alpha$ seems to be smaller than $\alpha_{\rm NN}=2$ for the NN RFIM. This is surprising because LR interactions in the corresponding pure system have the opposite effect of speeding up the growth process! In the present study, we were able to access time scales up to $t = 10^6$ MCS with system size $N = 2^{20}$. In order to arrive at more conclusive results, one should investigate the growth law up to (say) $t=10^8$ MCS at least. This would also require a considerably larger system size so that the results are not plagued by finite-size effects. 

It is useful to provide some details of the computational effort involved in these simulations. A single run up to $t = 10^6$ MCS on a system with $\sigma = 0.9$ and $\Delta = 0.05$ took approximately $247$ core-hours on the \textit{Intel Xeon Phi Processor 7210}. In order to reduce the computational time, we store the local field for each spin at the start of the simulation. Using this simple trick, we need to update the field only when a spin flip is accepted. This results in a significant speed-up, especially when the quench temperature is low and the disorder is large. To reach time-scales up to $t=10^8$ MCS for the current system sizes ($N = 2^{20}$), a $100$ times larger computational effort would be required. For larger systems, the required computational time would be further multiplied.

We also explored the superuniversality (SU) or disorder-independence in the correlation functions of the magnetization field, both at equal and separated times. For LR systems, the equal-time correlation function $C(r,t)$ exhibits dynamical scaling. In addition, the scaling function is independent of the disorder amplitude, i.e., the scaling function shows SU. We also discussed the SU of the autocorrelation function $A(t,t_{{\rm w}})$. We find that the scaling function for $A(t,t_{{\rm w}})$ exhibits SU for $\sigma < 1$, while for $\sigma > 1$ it appears to be violated. The latter case indicates behavior akin to the NN RFIM, where the violation of SU has been observed~\cite{PhysRevE.65.046114}.

For further insights, an analytical treatment of ordering dynamics of the $1D$ RFLRIM is desirable. Towards this goal, we have initiated a Langevin study of random walks with a systematic drift term in a Sinai potential. This is the LR counterpart of the free diffusion problem studied by Fisher et al.~\cite{PhysRevE.64.066107} for the $1D$ NN RFIM. Our understanding of this system will be presented elsewhere.

\appendix

{\section{Domain Growth Law from Spatial Correlation Function}
\label{corr_len}

The domain growth law can also be extracted from the decay of the spatial correlation function $C(r,t)$ as
\be
\label{crr}
C(r=R_c (t),t) = C_0,
\ee
where $C_0$ is a constant (say, $0.5$). Here, $R_c (t)$ serves as another measure of the coarsening length scale. In the scaling regime, $R_c (t)$ should be proportional to the length scale $R(t)$ obtained from the inverse defect density in the main text.

\begin{figure}[t!]
\centering
\includegraphics[width=0.85\linewidth]{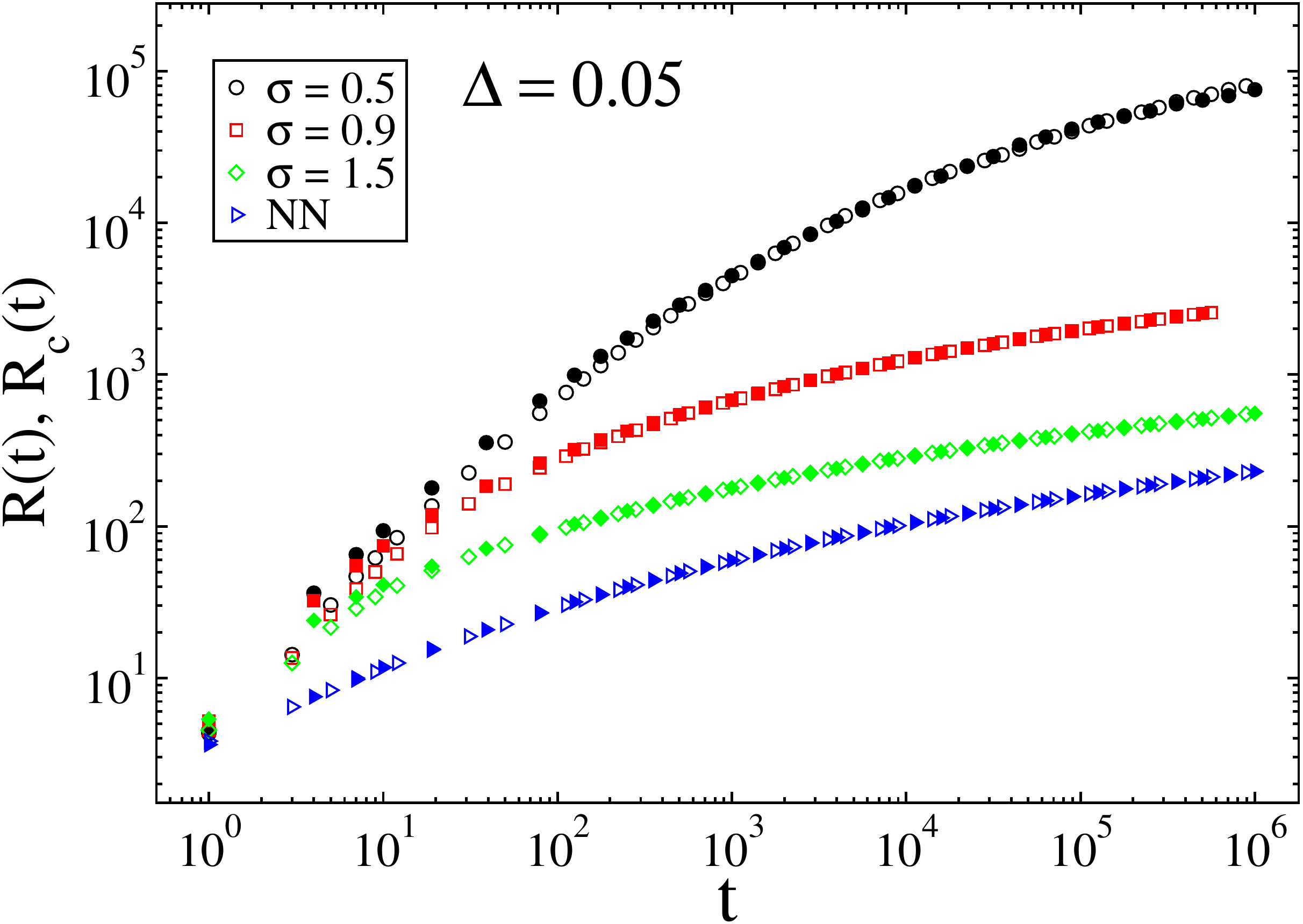}
\caption{Plot of length scales $R(t)$ and $R_c(t)$ vs. $t$ (on a log-log scale) for different $\sigma$ and the NN case, at fixed $\Delta = 0.05$. The empty symbols represent $R(t)$, while the filled symbols represent $R_c(t)$. The quantity $R_c(t)$ has been multiplied by an adjustable parameter for each set.}
\label{Rc_def}
\end{figure}

In Fig.~\ref{Rc_def}, we compare the quantities $R(t)$ and $R_c(t)$ for different $\sigma$ and the nearest-neighbor (NN) case, at a fixed $\Delta = 0.05$. We have scaled $R_c(t)$ by appropriate factors to fall on the data set for $R(t)$. For all $\sigma$ and the NN case, $R(t)$ and $R_c(t)$ are in excellent agreement.

\bibliography{rflrim}

\end{document}